\documentclass{article}

\PassOptionsToPackage{numbers, compress}{natbib}

\usepackage[final]{neurips_mlsb_2021}

\usepackage[utf8]{inputenc} % allow utf-8 input
\usepackage[T1]{fontenc}    % use 8-bit T1 fonts
\usepackage{hyperref}       % hyperlinks
\usepackage{url}            % simple URL typesetting
\usepackage{booktabs}       % professional-quality tables
\usepackage{amsfonts}       % blackboard math symbols
\usepackage{nicefrac}       % compact symbols for 1/2, etc.
\usepackage{microtype}      % microtypography

\usepackage{amsmath}
\newcommand{\angstrom}{\textup{\AA}}

\usepackage{pifont}% http://ctan.org/pkg/pifont
\newcommand{\cmark}{\ding{51}}%
\newcommand{\xmark}{\ding{55}}%
\usepackage{subcaption}
\usepackage{graphicx}
\renewcommand{\thefigure}{\arabic{figure}} 
\title{Learning physics confers pose-sensitivity in structure-based virtual screening}

\author{
  Pawel Gniewek\\
  Atomwise Inc\\
  San Francisco, CA, USA \\
  \texttt{pawel@atomwise.com} \\
  \And
  Bradley Worley \\
  Atomwise Inc\\
  San Francisco, CA, USA \\
  \texttt{brad@atomwise.com} \\
  \And
  Kate Stafford \\
  Atomwise Inc\\
  San Francisco, CA, USA \\
  \texttt{kate@atomwise.com} \\
  \And
  Henry van den Bedem \\
  Atomwise Inc \\
  University of California, San Francisco\\
  San Francisco, CA, USA \\
  \texttt{vdbedem@atomwise.com} \\
  \And
  Brandon Anderson \\
  Atomwise Inc\\
  San Francisco, CA, USA \\
  \texttt{brandon@atomwise.com} \\
}

\begin{document}

\maketitle

\begin{abstract}

In drug discovery, structure-based virtual high-throughput screening (vHTS) campaigns aim to identify bioactive ligands or ``hits'' for therapeutic protein targets from docked poses at specific binding sites. However, while generally successful at this task, many deep learning methods are known to be insensitive to protein-ligand interactions, decreasing the reliability of hit detection and hindering discovery at novel binding sites. Here, we overcome this limitation by introducing a class of models with two key features: 1) we condition bioactivity on pose quality score, and 2) we present poor poses of true binders to the model as negative examples. The conditioning forces the model to learn details of physical interactions. We evaluate these models on a new benchmark designed to detect pose-sensitivity.
\end{abstract}

\section{Introduction}
Vast, make-on-demand chemical libraries like ENAMINE or Mcule have transformed the scale of pharmaceutical, structure-based virtual high-throughput screening (vHTS) campaigns \cite{irwin_docking_2016}. To identify a ``hit'' from a library of candidate molecules, structure-based vHTS methods predict binding affinity between a protein and a ligand from their docked, bound complex, thereby assuming that experimentally observed affinities correlate with protein-ligand interactions. Conventional methods use empirical, physics-based approaches, which attempt to calculate the binding free energy of complex formation. By contrast newer deep learning (DL) approaches are trained on large data sets using implicit features to predict activity. These statistical models can outperform physics-based approaches in retrospective tests for predicting activity.

Early structure-based DL methods in vHTS centered on Convolutional Neural Networks (CNNs), representing protein-ligand structures by a 3D voxel grid to predict activity \cite{wallach_atomnet_2015,ragoza_proteinligand_2017, stepniewska-dziubinska_development_2018,boyles_learning_2019}. Although generally effective \cite{hsieh_miro1_2019}, a drawback of CNNs is that they are not rotationally invariant, and require more parameters than alternative representations. Consequently, Graph Convolutional Networks (GCNs)~\cite{kipf_semi-supervised_2017, gilmer_neural_2017}, and their position-based generalizations~\cite{behler_generalized_2007, schutt_schnet_2017, feinberg_potentialnet_2018, lim_predicting_2019, stafford_atomnet_2021, thomas_tensor_2018, anderson_cormorant_2019, townshend_atom3d_2021}, have gained popularity by capitalizing on physical symmetries and using data more efficiently.

Recent studies have suggested that the performance of structure-based machine learning methods is partly driven by proteochemometric (PCM)-like features \cite{boyles_learning_2019, sieg_need_2019, chen_hidden_2019}. Rather than responding to specific interactions between the ligand and the binding site, the model learns independent ligand and protein signatures. This deficiency manifests as a drop in predictive performance when the model is confronted with a previously unseen binding site on the same protein, especially when that site partially overlaps with a canonical site and portions of the canonical site lie within the receptive field of the network. Thus the model can be biased by detected but chemically irrelevant features provided in the input data -- analogous to the ``Picasso Problem'' in 2D CNNs. For example, the model may highly rank ATP-competitive binders for a neighboring allosteric site on a protein kinase. This limitation critically hinders discovery of new chemical matter, or the ability to target novel sites on proteins. 

Simultaneous training on ligand pose quality and affinity can improve pose sensitivity \cite{francoeur_three-dimensional_2020}. Here, we build on that observation and present a multi-task architecture which simultaneously evaluates bioactivity and the physics-based Vina \cite{trott_autodock_2010} score of the pose. We then condition its bioactivity task on a separate measure of pose quality, as defined by the score assigned to a pose by PoseRanker, a GCN developed to identify high-quality poses from docking \cite{stafford_atomnet_2021}. Finally, we use poor poses of true positive compounds as additional negative examples. We demonstrate that our architecture improves pose-sensitivity on several benchmarks.

\section{Methods}
\subsection{Neural Network Architectures}
Our model is a GCN-based architecture with atoms as vertices, and pair-wise, distance dependent edges. We consider only receptor atoms within 7\angstrom{} of any ligand atom. In the first two graph convolutional layers any two atoms share an edge if they are within 4\angstrom{} of each other. We then extract ligand-only features and follow with two ligand-only layers. We sum-pool the final ligand-only layer to create an embedding for the multi-task multilayer perceptrons at the top of the network. The network architecture is further described in Sections \ref{gcn_arch} and \ref{multitask}. We predict three outputs using these embeddings: bioactivity, the PoseRanker pose quality score, and the Vina docking score. This is performed in two stages. First, we compute the PoseRanker and Vina score prediction by passing the embedding through two independent multilayer perceptrons. We then form a conditioned embedding by concatenating the input embedding with the PoseRanker score prediction, which is passed to a third multilayer perceptron to compute the activity prediction \cite{long_conditional_2018}. Section \ref{model_training} describes model training parameters.

\begin{figure}
  \centering
  \includegraphics[width=10cm]{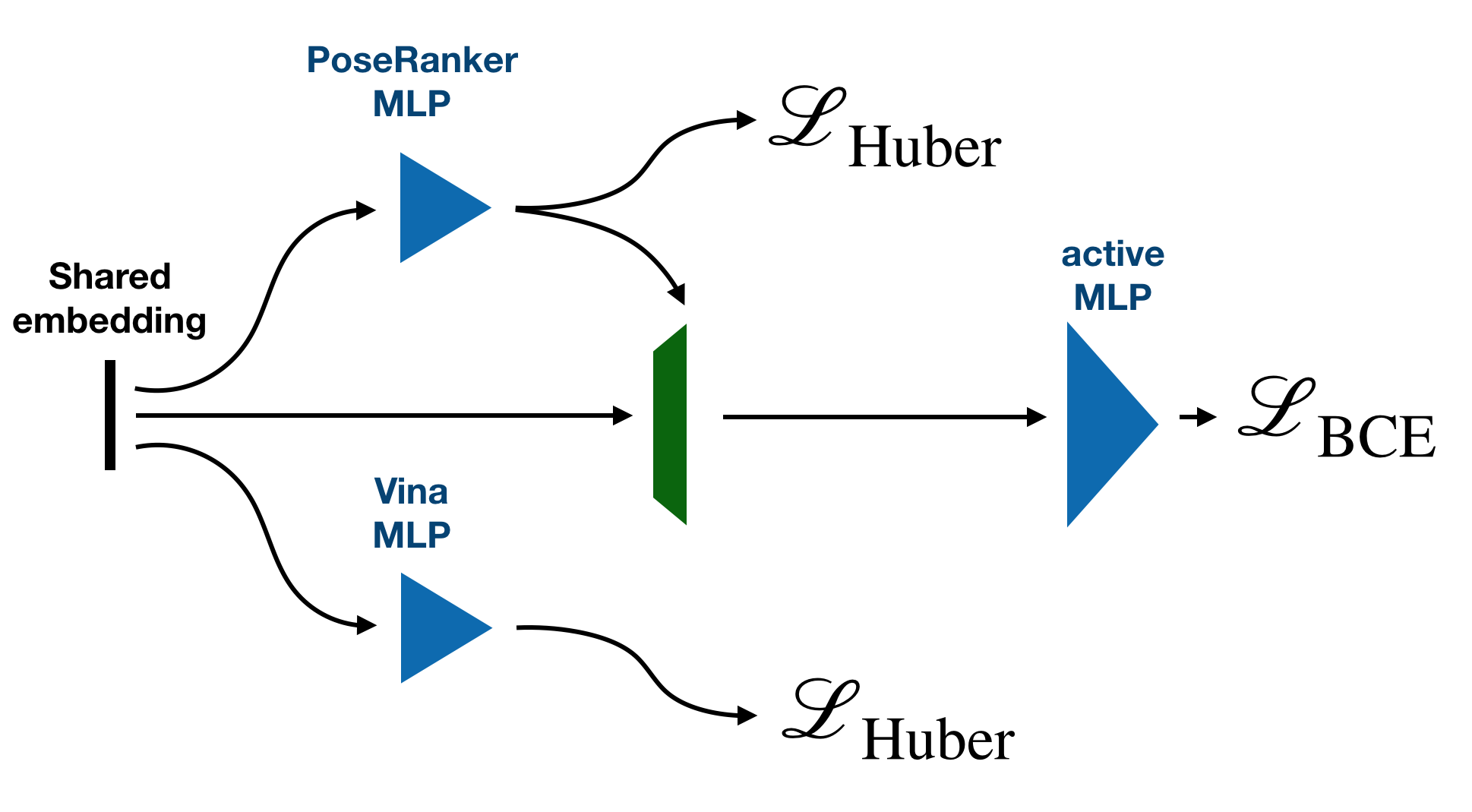}
  \caption{Multi-task architecture of the MT-4b model.}
  \label{mt}
\end{figure}

\begin{table}
  \caption{Deep Learning Neural Networks used in this work. Architectures are either a 3D CNN-based model or a GCN-based model. Multi-task (MT) models include additional Vina \cite{trott_autodock_2010,morrison_cuina_2020} or PoseRanker \cite{stafford_atomnet_2021} tasks, and may include additional conditioning layer. }
  \label{sample-table}
  \centering
  \begin{tabular}{llcccc}
    \toprule
    Name & Architecture & Vina & PoseRanker & Pose-negatives & Cond. \\
    \midrule
    CNN & 3D CNN based & \xmark &\xmark &\xmark& \xmark\\
    GCN & GCN based & \xmark & \xmark & \xmark & \xmark \\
    MT-1 & GCN based & \cmark & \xmark & \xmark & \xmark \\
    MT-2 & GCN based & \cmark & \cmark & \xmark & \xmark \\
    MT-3 & GCN based & \cmark & \cmark & \xmark & \cmark \\
    MT-4a & GCN based & \cmark & \cmark & \cmark & \xmark \\
    MT-4b & GCN based & \cmark & \cmark & \cmark & \cmark \\
    \bottomrule
  \end{tabular}
  \label{tab:mt}
\end{table}

\subsection{Data}
The training dataset consists of binding affinity measurements collected from publicly available sources (Chembl, Pubchem) and commercial databases (Reaxys, Liceptor). We considered quantitative measurements only with $\textrm{pKi} \in (0, 11)$. We labeled compounds as active if their measured pKi (or pIC50) is less than 10$\mu$M; otherwise they were labeled as inactive. 
There are more active than inactive compounds, so to balance the training set we randomly assigned each of the active compounds as inactive decoys for another, dissimilar protein target. For some models we also used pose-negatives - poor poses of active compounds treated as negatives (Section \ref{pns}). We selected a set of 12 diverse proteins (D12) to serve as a hold-out test set and excluded all close homologs of the D12 proteins (>95\% sequence similarity) from training. Training set covers more than 3800 diverse proteins, and counts 4.8M (5.8M) data points without (with) pose-negatives. The hold-out set counts about 33,000 compounds and 12 proteins. We docked compounds with the GPU-enabled docking engine, CUina \cite{morrison_cuina_2020}. We used the PoseRanker \cite{stafford_atomnet_2021} top-ranked pose for scoring.

\subsection{Numerical experiments}
To evaluate pose-sensitivity of our models, we scored each of the active target-compound pairs in D12 three times using: i) the top pose, ii) a poor pose, iii) the average score of four physically implausible poses, each obtained by randomly rotating the top pose around its center of mass. (See Figure 2a.) Our measure of pose-sensitivity is the median of the drop of the bioactivity score between good and poor/implausible poses. We expect that a model sensitive to ligand-receptor interactions reveals sharply reduced bioactivity scores for poor or implausible poses compared to good poses.

\begin{figure}
  \begin{subfigure}{7cm}
    \centering\includegraphics[width=7cm]{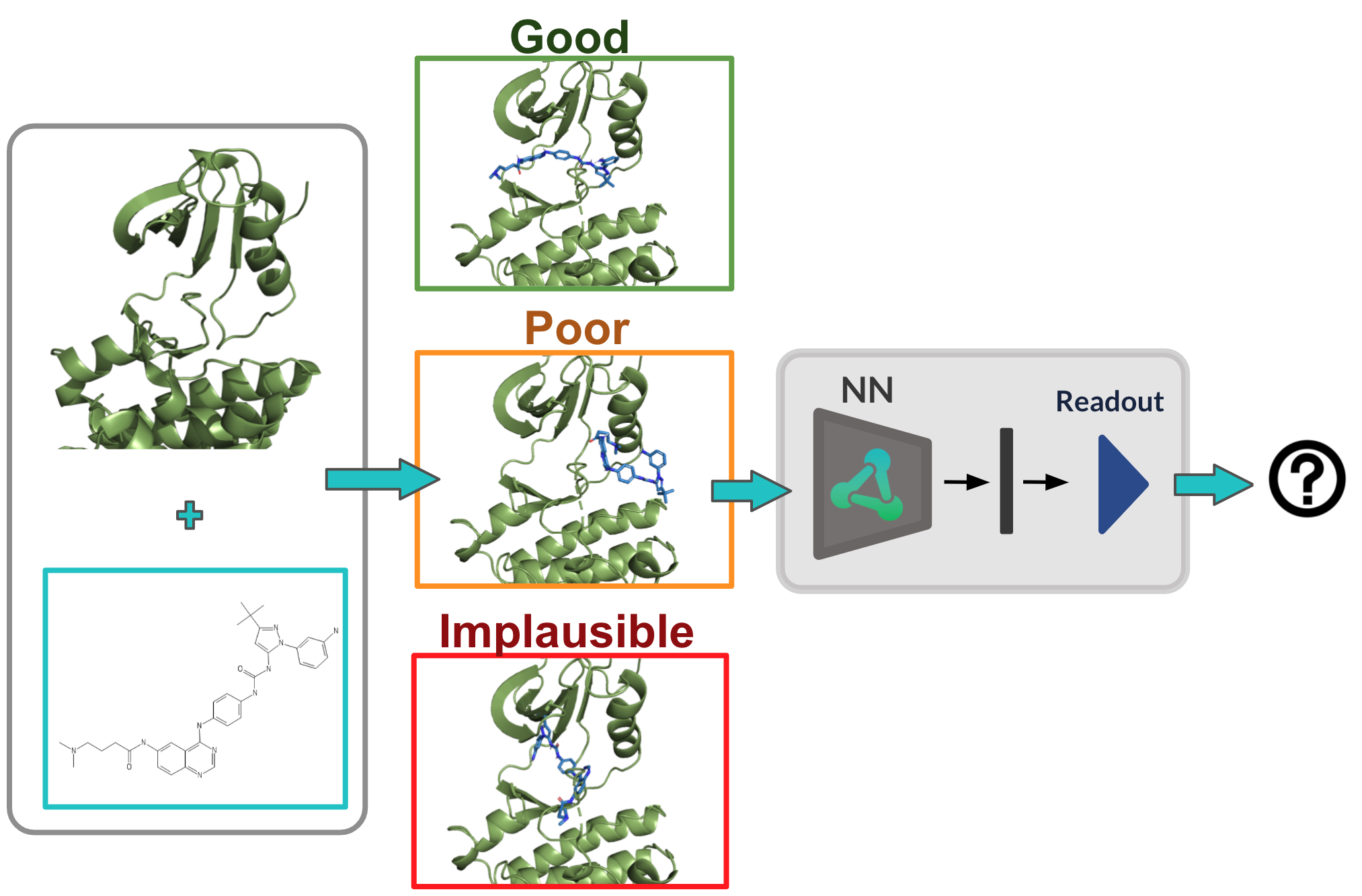}
  \end{subfigure}
  \begin{subfigure}{7cm}
    \centering\includegraphics[width=7cm]{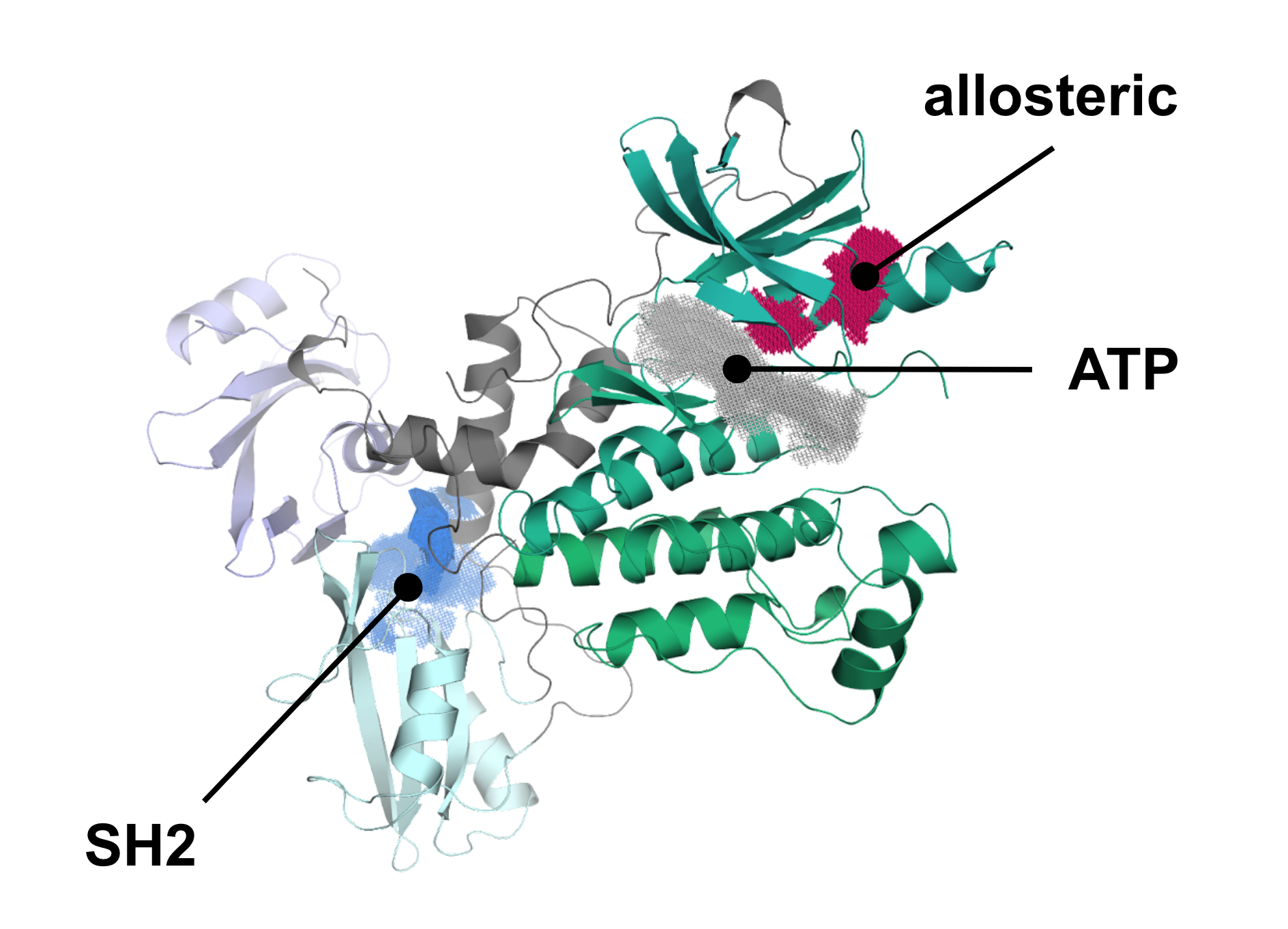}
  \end{subfigure}
  \caption{(\textbf{a}) A schematic of the pose-sensitivity experiment. For each target-ligand pair we generate three types of poses: i) a good pose ii) a poor pose and iii) an implausible pose. Good and poor poses are the highest and lowest ranked poses by PoseRanker after sampling with CUina. Implausible poses are obtained from good poses by a random rotation of the ligand around its center of mass. Scores for the poor and implausible poses are subtracted from the score of the good pose. (\textbf{b}) Human ZAP 70 protein (PDB: 2ozo) indicating ATP site (grey), allosteric site (red), and spatially distant SH2 site (blue). A set of $10^5$ compounds is docked to each site and scored with the Deep Learning models.}
  \label{poses_zap70}
\end{figure}

Next, we designed a benchmark based on a protein with multiple binding sites to investigate the Picasso Problem described above. To monitor how the neighbor binding sites interfere with the model’s inference, we selected three binding sites on the same protein kinase hZAP70: the ATP site, a nearby allosteric site 6-10{\AA} away, and a distant SH2 site over 50{\AA} away (Figure \ref{poses_zap70}). We selected a diverse set of known ATP-site-binding kinase inhibitors (about 300 compounds labeled as actives) and mixed them with $10^5$ randomly selected compounds from a screening library (MCULE (17/08/2018); labeled as inactive). Each compound was docked to each site. We expect that a good model can distinguish kinase inhibitors against a background of random molecules docked to the ATP binding site (ROC AUC $\gg$ 0.5). However, a pose-sensitive model should not be biased by the neighbor ATP-site when compounds are docked to the allosteric site, and therefore should have near random performance. To account for possible biases in the training set, we also examined the distant control site on the SH2 domain, which likewise does not bind ATP-site targeted molecules and is expected to have near-random performance in identifying ATP-site binders.

\section{Results}
\begin{figure}
 \centering
  \includegraphics[width=13.5cm]{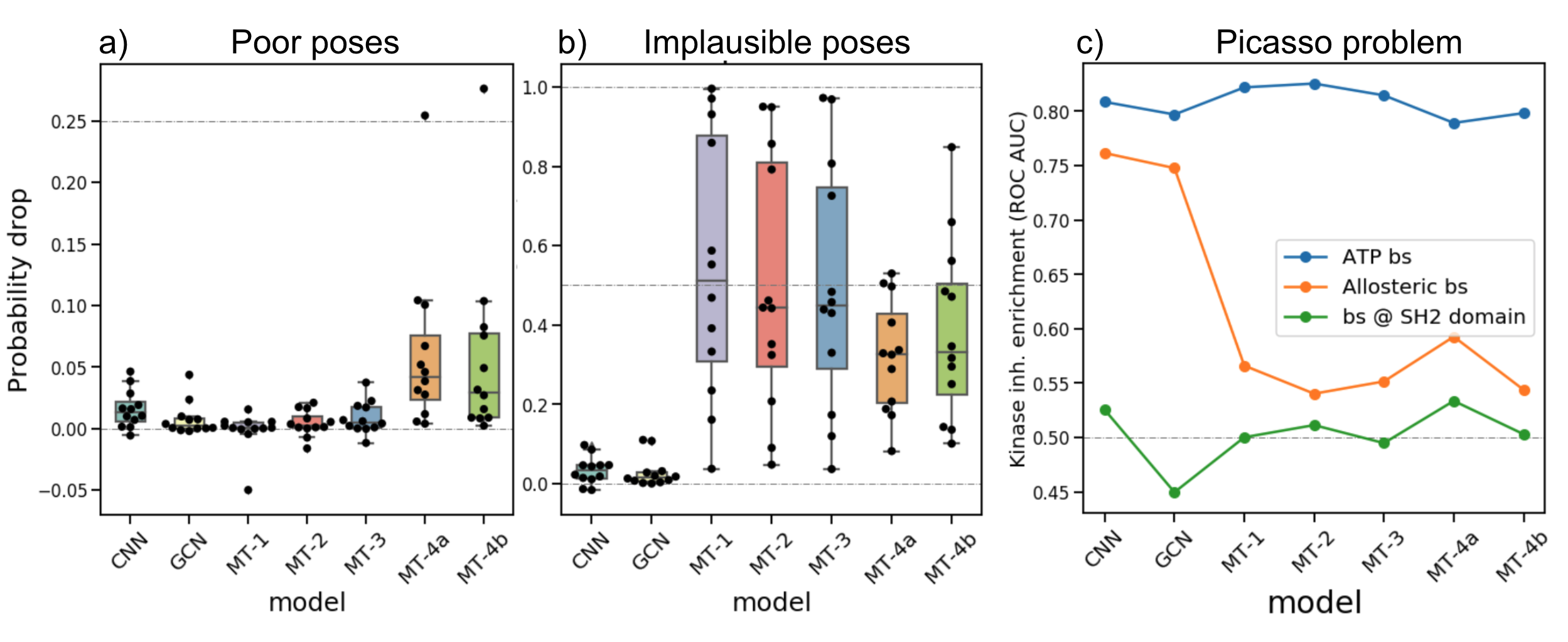}

\caption{(\textbf{a}) Median probability drops between good and poor poses. (\textbf{b}) Median probability drops between good and implausible poses. (\textbf{c}) Picasso problem experiment. $10^5$ diverse compounds (labeled as 0, non-binders) mixed with {\it c.a.} 300 kinase inhibitors (labeled as 1, binders) were docked and scored with at three binding sites i) ATP binding site ii) allosteric binding site iii) binding site at the SH2 domain, See Figure \ref{poses_zap70}b.}
  \label{drops}
\end{figure}

Figure \ref{d12} shows that our baseline models perform well on the holdout set, with GCN-based models slightly outperforming CNN. However, both single-task models enrich ATP-site compounds at the allosteric site (Figure \ref{drops}). Because the ATP site is in the receptive field of these two networks when evaluating the allosteric site, GCN and CNN models make predictions as if they were shown the ATP site (Figure \ref{poses_zap70}b). This cannot be due to a biased training set, as screening the spatially distant SH2 site gave no enrichment of ATP-site compounds (Figure \ref{drops}c). In other words, the single-task CNN and GCN models are effectively PCMs. Further, they are insensitive not only to poor poses, but also to implausible poses (Figure \ref{drops}a,b). This behavior was observed previously with 3D-grid-based CNNs \cite{stepniewska-dziubinska_development_2018,francoeur_three-dimensional_2020}, but a general solution has been lacking. 

At first it is unintuitive that the single task activity models trained on structural data do not use that structural information about ligand-receptor interactions. Apparently, the model has no direct incentive to learn the structural features of the ligand-receptor interactions, so models often neglect them. Training a multi-task model where the additional tasks require embeddings sensitive to structure should alleviate the problem. Indeed, the Multi-Task (MT) models show increased pose-sensitivity. Adding a docking score regression task (model MT-1) leads to a model that penalizes implausible poses and decreases the enrichment of kinase inhibitors at the allosteric site of the hZAP70 protein (Figure \ref{drops}b,c). However, because poor poses are generally far from native but do not have any atomic clashes, the MT-1 model is unable to distinguish between good and poor poses (Figure \ref{drops}a). Interestingly, adding pose quality regression as the third task (MT-2) or conditioning the activity task on the pose quality (MT-3) do not increase sensitivity to poor poses. Models shown only top-ranked poses do not learn what bad poses look like. To correct for this, we introduced "pose negatives": bottom-ranked poses of true actives treated as negative examples. The resulting models (MT-4a and MT-4b) penalize physically implausible poses as well as poor poses. The same models also mitigate the Picasso problem. The model that does not condition its activity task on pose quality (MT-4a) is still somewhat more prone to the Picasso problem compared to the conditioning model (MT-4b).

\section{Conclusions}
Multi-task architectures can lead to models that are capable of predicting bioactivity of compounds while making full use of the structural data provided for the inference. Forcing the model to learn orthogonal tasks regularizes the final model. Pose sensitive models enable structure-based deep learning approaches in drug discovery to target novel binding sites, and unlock previously undruggable proteins.

%\newpage

%\section*{Broader Impact}
%The present work was developed in the context of a drug discovery pipeline designed to reduce the costs and development time associated with early-stage drug discovery. Success in this area can improve access to medication and reduce health-care costs.

%%THIS is not really broader impact, but rather 
%%Note that our training dataset consists largely of publicly available data and thus necessarily reflects biases in allocation of research funding to various diseases and health conditions. We anticipate that improving the generalizability of our model across novel protein targets and binding sites can help mitigate this limitation in the training data.

\begin{ack}
We thank Michael Mysinger and Adrian Stecula for helpful comments.
\end{ack}

%\section*{References}
%\medskip

\bibliography{references}

\begin{thebibliography}{22}
\providecommand{\natexlab}[1]{#1}
\providecommand{\url}[1]{\texttt{#1}}
\expandafter\ifx\csname urlstyle\endcsname\relax
  \providecommand{\doi}[1]{doi: #1}\else
  \providecommand{\doi}{doi: \begingroup \urlstyle{rm}\Url}\fi

\bibitem[Irwin and Shoichet(2016)]{irwin_docking_2016}
John~J. Irwin and Brian~K. Shoichet.
\newblock Docking {Screens} for {Novel} {Ligands} {Conferring} {New} {Biology}:
  {Miniperspective}.
\newblock \emph{Journal of Medicinal Chemistry}, 59\penalty0 (9):\penalty0
  4103--4120, May 2016.
\newblock ISSN 0022-2623, 1520-4804.
\newblock \doi{10.1021/acs.jmedchem.5b02008}.
\newblock URL \url{https://pubs.acs.org/doi/10.1021/acs.jmedchem.5b02008}.

\bibitem[Wallach et~al.(2015)Wallach, Dzamba, and
  Heifets]{wallach_atomnet_2015}
Izhar Wallach, Michael Dzamba, and Abraham Heifets.
\newblock {AtomNet}: {A} {Deep} {Convolutional} {Neural} {Network} for
  {Bioactivity} {Prediction} in {Structure}-based {Drug} {Discovery}.
\newblock \emph{arXiv:1510.02855 [cs, q-bio, stat]}, October 2015.
\newblock URL \url{http://arxiv.org/abs/1510.02855}.
\newblock arXiv: 1510.02855.

\bibitem[Ragoza et~al.(2017)Ragoza, Hochuli, Idrobo, Sunseri, and
  Koes]{ragoza_proteinligand_2017}
Matthew Ragoza, Joshua Hochuli, Elisa Idrobo, Jocelyn Sunseri, and David~Ryan
  Koes.
\newblock Protein–{Ligand} {Scoring} with {Convolutional} {Neural}
  {Networks}.
\newblock \emph{Journal of Chemical Information and Modeling}, 57\penalty0
  (4):\penalty0 942--957, April 2017.
\newblock ISSN 1549-9596, 1549-960X.
\newblock \doi{10.1021/acs.jcim.6b00740}.
\newblock URL \url{https://pubs.acs.org/doi/10.1021/acs.jcim.6b00740}.

\bibitem[Stepniewska-Dziubinska et~al.(2018)Stepniewska-Dziubinska,
  Zielenkiewicz, and Siedlecki]{stepniewska-dziubinska_development_2018}
Marta~M Stepniewska-Dziubinska, Piotr Zielenkiewicz, and Pawel Siedlecki.
\newblock Development and evaluation of a deep learning model for
  protein–ligand binding affinity prediction.
\newblock \emph{Bioinformatics}, 34\penalty0 (21):\penalty0 3666--3674,
  November 2018.
\newblock ISSN 1367-4803, 1460-2059.
\newblock \doi{10.1093/bioinformatics/bty374}.
\newblock URL
  \url{https://academic.oup.com/bioinformatics/article/34/21/3666/4994792}.

\bibitem[Boyles et~al.(2019)Boyles, Deane, and Morris]{boyles_learning_2019}
Fergus Boyles, Charlotte~M Deane, and Garrett~M Morris.
\newblock Learning from the ligand: using ligand-based features to improve
  binding affinity prediction.
\newblock \emph{Bioinformatics}, page btz665, August 2019.
\newblock ISSN 1367-4803, 1460-2059.
\newblock \doi{10.1093/bioinformatics/btz665}.
\newblock URL
  \url{https://academic.oup.com/bioinformatics/advance-article/doi/10.1093/bioinformatics/btz665/5554651}.

\bibitem[Hsieh et~al.(2019)Hsieh, Li, Vanhauwaert, Nguyen, Davis, Bu, Wszolek,
  and Wang]{hsieh_miro1_2019}
Chung-Han Hsieh, Li~Li, Roeland Vanhauwaert, Kong~T. Nguyen, Mary~D. Davis,
  Guojun Bu, Zbigniew~K. Wszolek, and Xinnan Wang.
\newblock Miro1 {Marks} {Parkinson}’s {Disease} {Subset} and {Miro1}
  {Reducer} {Rescues} {Neuron} {Loss} in {Parkinson}’s {Models}.
\newblock \emph{Cell Metabolism}, 30\penalty0 (6):\penalty0 1131--1140.e7,
  December 2019.
\newblock ISSN 15504131.
\newblock \doi{10.1016/j.cmet.2019.08.023}.
\newblock URL
  \url{https://linkinghub.elsevier.com/retrieve/pii/S1550413119304978}.

\bibitem[Kipf and Welling(2017)]{kipf_semi-supervised_2017}
Thomas~N. Kipf and Max Welling.
\newblock Semi-{Supervised} {Classification} with {Graph} {Convolutional}
  {Networks}.
\newblock \emph{arXiv:1609.02907 [cs, stat]}, February 2017.
\newblock URL \url{http://arxiv.org/abs/1609.02907}.
\newblock arXiv: 1609.02907.

\bibitem[Gilmer et~al.(2017)Gilmer, Schoenholz, Riley, Vinyals, and
  Dahl]{gilmer_neural_2017}
Justin Gilmer, Samuel~S. Schoenholz, Patrick~F. Riley, Oriol Vinyals, and
  George~E. Dahl.
\newblock Neural message passing for {Quantum} chemistry.
\newblock In \emph{Proceedings of the 34th {International} {Conference} on
  {Machine} {Learning} - {Volume} 70}, {ICML}'17, pages 1263--1272, Sydney,
  NSW, Australia, August 2017. JMLR.org.

\bibitem[Behler and Parrinello(2007)]{behler_generalized_2007}
Jörg Behler and Michele Parrinello.
\newblock Generalized {Neural}-{Network} {Representation} of
  {High}-{Dimensional} {Potential}-{Energy} {Surfaces}.
\newblock \emph{Physical Review Letters}, 98\penalty0 (14):\penalty0 146401,
  April 2007.
\newblock \doi{10.1103/PhysRevLett.98.146401}.
\newblock URL \url{https://link.aps.org/doi/10.1103/PhysRevLett.98.146401}.

\bibitem[Schütt et~al.(2017)Schütt, Kindermans, Sauceda, Chmiela, Tkatchenko,
  and Müller]{schutt_schnet_2017}
Kristof~T. Schütt, Pieter-Jan Kindermans, Huziel~E. Sauceda, Stefan Chmiela,
  Alexandre Tkatchenko, and Klaus-Robert Müller.
\newblock {SchNet}: {A} continuous-filter convolutional neural network for
  modeling quantum interactions.
\newblock \emph{arXiv:1706.08566 [physics, stat]}, December 2017.
\newblock URL \url{http://arxiv.org/abs/1706.08566}.
\newblock arXiv: 1706.08566.

\bibitem[Feinberg et~al.(2018)Feinberg, Sur, Wu, Husic, Mai, Li, Sun, Yang,
  Ramsundar, and Pande]{feinberg_potentialnet_2018}
Evan~N. Feinberg, Debnil Sur, Zhenqin Wu, Brooke~E. Husic, Huanghao Mai, Yang
  Li, Saisai Sun, Jianyi Yang, Bharath Ramsundar, and Vijay~S. Pande.
\newblock {PotentialNet} for {Molecular} {Property} {Prediction}.
\newblock \emph{ACS Central Science}, 4\penalty0 (11):\penalty0 1520--1530,
  November 2018.
\newblock ISSN 2374-7943, 2374-7951.
\newblock \doi{10.1021/acscentsci.8b00507}.
\newblock URL \url{https://pubs.acs.org/doi/10.1021/acscentsci.8b00507}.

\bibitem[Lim et~al.(2019)Lim, Ryu, Park, Choe, Ham, and
  Kim]{lim_predicting_2019}
Jaechang Lim, Seongok Ryu, Kyubyong Park, Yo~Joong Choe, Jiyeon Ham, and
  Woo~Youn Kim.
\newblock Predicting {Drug}–{Target} {Interaction} {Using} a {Novel} {Graph}
  {Neural} {Network} with {3D} {Structure}-{Embedded} {Graph} {Representation}.
\newblock \emph{Journal of Chemical Information and Modeling}, 59\penalty0
  (9):\penalty0 3981--3988, September 2019.
\newblock ISSN 1549-9596, 1549-960X.
\newblock \doi{10.1021/acs.jcim.9b00387}.
\newblock URL \url{https://pubs.acs.org/doi/10.1021/acs.jcim.9b00387}.

\bibitem[Stafford et~al.(2021)Stafford, Anderson, Sorenson, and
  Bedem]{stafford_atomnet_2021}
Kate Stafford, Brandon~M. Anderson, Jon Sorenson, and Henry van~den Bedem.
\newblock {AtomNet} {PoseRanker}: {Enriching} {Ligand} {Pose} {Quality} for
  {Dynamic} {Proteins} in {Virtual} {High} {Throughput} {Screens}.
\newblock September 2021.
\newblock \doi{10.33774/chemrxiv-2021-t6xkj}.
\newblock URL
  \url{https://chemrxiv.org/engage/chemrxiv/article-details/614b905e39ef6a1c36268003}.

\bibitem[Thomas et~al.(2018)Thomas, Smidt, Kearnes, Yang, Li, Kohlhoff, and
  Riley]{thomas_tensor_2018}
Nathaniel Thomas, Tess Smidt, Steven Kearnes, Lusann Yang, Li~Li, Kai Kohlhoff,
  and Patrick Riley.
\newblock Tensor field networks: {Rotation}- and translation-equivariant neural
  networks for {3D} point clouds.
\newblock \emph{arXiv:1802.08219 [cs]}, May 2018.
\newblock URL \url{http://arxiv.org/abs/1802.08219}.
\newblock arXiv: 1802.08219.

\bibitem[Anderson et~al.(2019)Anderson, Hy, and
  Kondor]{anderson_cormorant_2019}
Brandon Anderson, Truong~Son Hy, and Risi Kondor.
\newblock Cormorant: {Covariant} {Molecular} {Neural} {Networks}.
\newblock In \emph{Advances in {Neural} {Information} {Processing} {Systems}},
  volume~32. Curran Associates, Inc., 2019.
\newblock URL
  \url{https://papers.nips.cc/paper/2019/hash/03573b32b2746e6e8ca98b9123f2249b-Abstract.html}.

\bibitem[Townshend et~al.(2021)Townshend, Vögele, Suriana, Derry, Powers,
  Laloudakis, Balachandar, Jing, Anderson, Eismann, Kondor, Altman, and
  Dror]{townshend_atom3d_2021}
Raphael J.~L. Townshend, Martin Vögele, Patricia Suriana, Alexander Derry,
  Alexander Powers, Yianni Laloudakis, Sidhika Balachandar, Bowen Jing, Brandon
  Anderson, Stephan Eismann, Risi Kondor, Russ~B. Altman, and Ron~O. Dror.
\newblock {ATOM3D}: {Tasks} {On} {Molecules} in {Three} {Dimensions}.
\newblock \emph{arXiv:2012.04035 [physics, q-bio]}, June 2021.
\newblock URL \url{http://arxiv.org/abs/2012.04035}.
\newblock arXiv: 2012.04035.

\bibitem[Sieg et~al.(2019)Sieg, Flachsenberg, and Rarey]{sieg_need_2019}
Jochen Sieg, Florian Flachsenberg, and Matthias Rarey.
\newblock In {Need} of {Bias} {Control}: {Evaluating} {Chemical} {Data} for
  {Machine} {Learning} in {Structure}-{Based} {Virtual} {Screening}.
\newblock \emph{Journal of Chemical Information and Modeling}, 59\penalty0
  (3):\penalty0 947--961, March 2019.
\newblock ISSN 1549-9596, 1549-960X.
\newblock \doi{10.1021/acs.jcim.8b00712}.
\newblock URL \url{https://pubs.acs.org/doi/10.1021/acs.jcim.8b00712}.

\bibitem[Chen et~al.(2019)Chen, Cruz, Ramsey, Dickson, Duca, Hornak, Koes, and
  Kurtzman]{chen_hidden_2019}
Lieyang Chen, Anthony Cruz, Steven Ramsey, Callum~J. Dickson, Jose~S. Duca,
  Viktor Hornak, David~R. Koes, and Tom Kurtzman.
\newblock Hidden bias in the {DUD}-{E} dataset leads to misleading performance
  of deep learning in structure-based virtual screening.
\newblock \emph{PLOS ONE}, 14\penalty0 (8):\penalty0 e0220113, August 2019.
\newblock ISSN 1932-6203.
\newblock \doi{10.1371/journal.pone.0220113}.
\newblock URL \url{https://dx.plos.org/10.1371/journal.pone.0220113}.

\bibitem[Francoeur et~al.(2020)Francoeur, Masuda, Sunseri, Jia, Iovanisci,
  Snyder, and Koes]{francoeur_three-dimensional_2020}
Paul~G. Francoeur, Tomohide Masuda, Jocelyn Sunseri, Andrew Jia, Richard~B.
  Iovanisci, Ian Snyder, and David~R. Koes.
\newblock Three-{Dimensional} {Convolutional} {Neural} {Networks} and a
  {Cross}-{Docked} {Data} {Set} for {Structure}-{Based} {Drug} {Design}.
\newblock \emph{Journal of Chemical Information and Modeling}, 60\penalty0
  (9):\penalty0 4200--4215, September 2020.
\newblock ISSN 1549-9596, 1549-960X.
\newblock \doi{10.1021/acs.jcim.0c00411}.
\newblock URL \url{https://pubs.acs.org/doi/10.1021/acs.jcim.0c00411}.

\bibitem[Trott and Olson(2010)]{trott_autodock_2010}
Oleg Trott and Arthur~J. Olson.
\newblock {AutoDock} {Vina}: {Improving} the speed and accuracy of docking with
  a new scoring function, efficient optimization, and multithreading.
\newblock \emph{Journal of Computational Chemistry}, 31\penalty0 (2):\penalty0
  455--461, 2010.
\newblock ISSN 1096-987X.
\newblock \doi{https://doi.org/10.1002/jcc.21334}.
\newblock URL \url{https://onlinelibrary.wiley.com/doi/abs/10.1002/jcc.21334}.
\newblock \_eprint: https://onlinelibrary.wiley.com/doi/pdf/10.1002/jcc.21334.

\bibitem[Long et~al.(2018)Long, Cao, Wang, and Jordan]{long_conditional_2018}
Mingsheng Long, Zhangjie Cao, Jianmin Wang, and Michael~I. Jordan.
\newblock Conditional {Adversarial} {Domain} {Adaptation}.
\newblock \emph{arXiv:1705.10667 [cs]}, December 2018.
\newblock URL \url{http://arxiv.org/abs/1705.10667}.
\newblock arXiv: 1705.10667.

\bibitem[Morrison et~al.(2020)Morrison, Friedland, and
  Wallach]{morrison_cuina_2020}
Adrian Morrison, Greg Friedland, and Izhar Wallach.
\newblock {CUina}: {An} {Efficient} {GPU} {Implementation} of {AutoDock}
  {Vina}.
\newblock August 2020.
\newblock URL
  \url{https://blog.atomwise.com/efficient-gpu-implementation-of-autodock-vina}.

\end{thebibliography}

\small

\appendix
\section{Appendix}
\setcounter{figure}{0}  % resets figure numbers
\renewcommand{\thefigure} {S\arabic{figure}}    % changes figure labels to S1, S2 ...

\subsection{Graph Neural Network architecture}\label{gcn_arch}
\begin{figure}
  \centering
  \includegraphics[width=13.5cm]{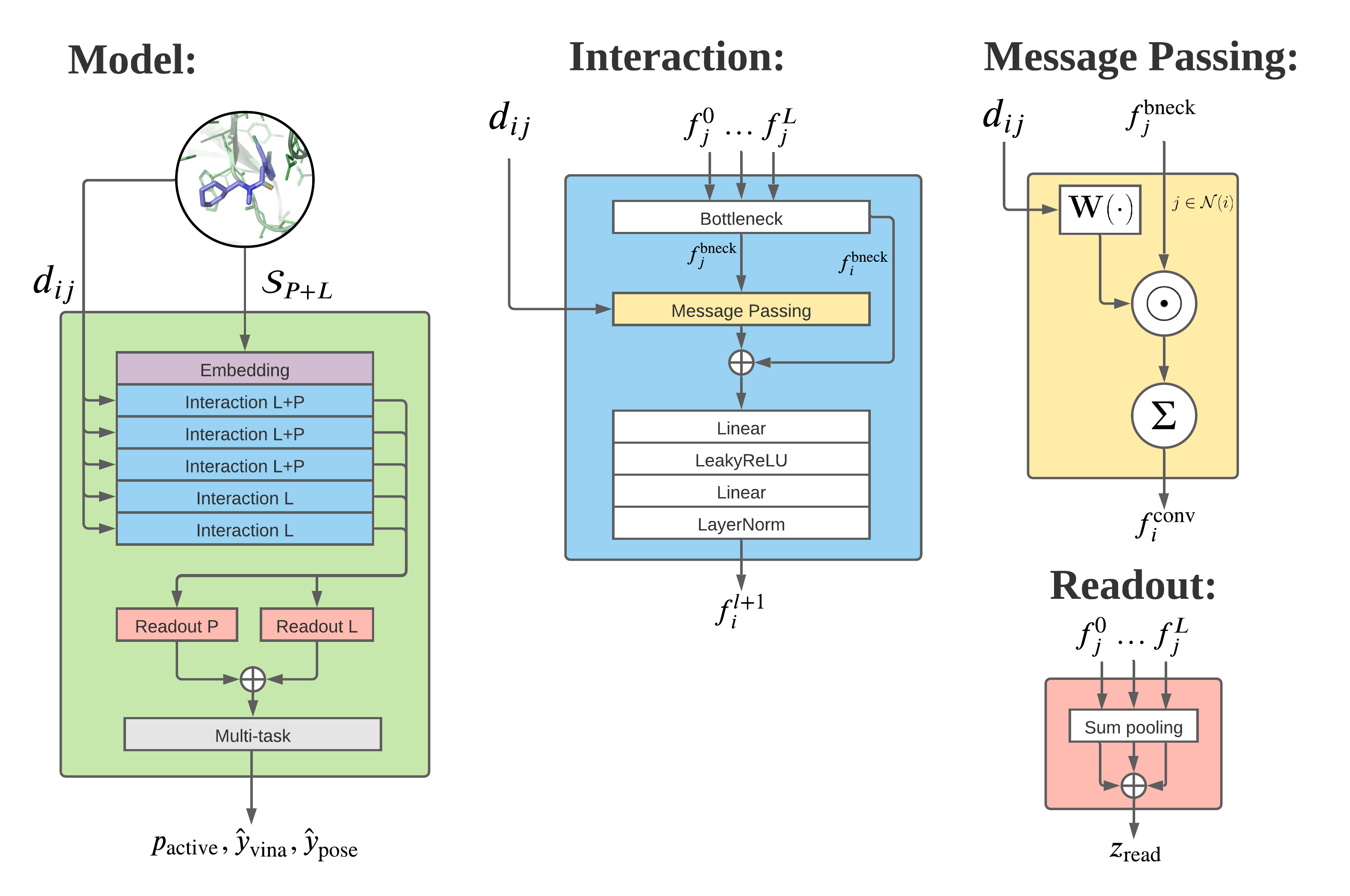}
  \caption{Architecture of the Graph Cronvolutional Network.}
  \label{fig:GCN_arch}
\end{figure}

Throughout the paper we use a graph convolutional network (GCN) to generate the embeddings used in our conditional multi-tasking architecture. Our GCN takes the three dimensional coordinates of the protein-ligand pose, along with a one-hot atom encoding that simultaneously identifies element type, protein/ligand, and hybridization state. We define our connectivity purely by radial functions without use of chemical bonds. See Figure~\ref{fig:GCN_arch} for details.

We use a radial cutoff $R^l_c$ at layer $l$ to define a radial graph, with the neighborhood for atom $i$ defined as $\mathcal{N}(i) = \{j : d_{ij} < R_c \}$, where $d_{ij} = \| \mathbf{r}_i - \mathbf{r}_j \|$ is the pairwise distance between atoms $i$ and $j$. At each layer $l$ we have a feature vector $f^l_i \in \mathcal{R}^{N_f^l}$ at atom $i \in \mathcal{S}_{P/L/P+L}$, where $N_f^l$ is the number of features at layer $l$, and $\mathcal{S}_{P}$, $\mathcal{S}_{L}$,  $\mathcal{S}_{P+L} = \mathcal{S}_{P} \cup \mathcal{S}_{L}$ are the set of atoms in the protein, ligand, or protein-ligand complex respectively. Our network can be configured so that $P$, $L$, or $P+L$ atoms can be used as either source atoms ($i$) or target atoms ($j$) on a layer-by-layer basis.

Our graph convolutional block is based upon a continuous filter convolution~\cite{schutt_schnet_2017}, $f^{\rm conv}_i \leftarrow \sum_j W^l(d_{ij}) f^l_j$. We use convolutional kernels $W^l(r) = \sum_{n}^{N_b^l} w_{n}^l j_0(r z_{0n} / R^l_c)$ constructed from a linear combinations of zeroth-order spherical Bessel functions $j_0(x) = \sin x / x$. Here $z_{0n}$ is the $n$-th zero of $j_0(x)$, and $w_n^l$ are learnable weights, and the number of basis functions is chosen as $N_b^l = \lceil R_c^l / \delta_l \rceil$, where $\delta_l = 0.15625 \angstrom$. This parametrization ensures that the convolutional kernels vanish at the edge of an internal neighborhood. Since we are not calculating forces, we do not need a smooth cutoff here. After the graph convolution step, we include a ${\rm Linear} - {\rm LeakyReLU} - {\rm Linear}$ layer. We do not apply a residual connection at the output of a convolution block. Instead, each convolutional layer takes as input all layers using a bottleneck layer. We found this performed better empirically than standard skip connections from the previous layer of the same set of source atoms.

Our network has five total graph convolutional blocks. In the first two graph conv blocks, we consider all ligand and receptor atoms so $i, j \in S_{P+L}$, and choose $R_c = 5\angstrom$ and $N^l_F = 64$ filters. The third graph conv block, the cutoff radius and filters are increased to $R_c = 7 \angstrom$ and $N^l_F = 128$, however only atoms in the ligand are considered as destination atoms (i. e., $i \in S_{P+L}$, but $j \in S_L$ only). Finally, in the last two layers only ligand atom embeddings are aggregated ($i, j \in S_{L}$) with a cutoff $R_c = 7\angstrom$ and $ N^l_F = 128$.

To construct our final embedding, we first apply a readout operation to the protein and ligand features independently at each layer: $z_P^l = \sum_{i \in S_P} f_i^l$, $z_L^l =  \sum_{i \in S_L} f_i^l$. We then concatenate these embeddings together, ${\bf z}_{\rm read} = \bigoplus_{l=0}^1 z_P^l \oplus \bigoplus_{l=0}^5 z_L^l$, and finally construct ${\bf z}$ from two iterations of ${\rm Dropout}_{0.2} - {\rm LeakyReLU} - {\rm Linear}$ applied to ${\bf z}_{\rm read}$. This final embedding is then used as the input to the conditional multi-task prediction layer.

\subsection{Conditional Multi-task architectures}
\label{multitask}
\begin{figure}
  \centering
  \includegraphics[width=10cm]{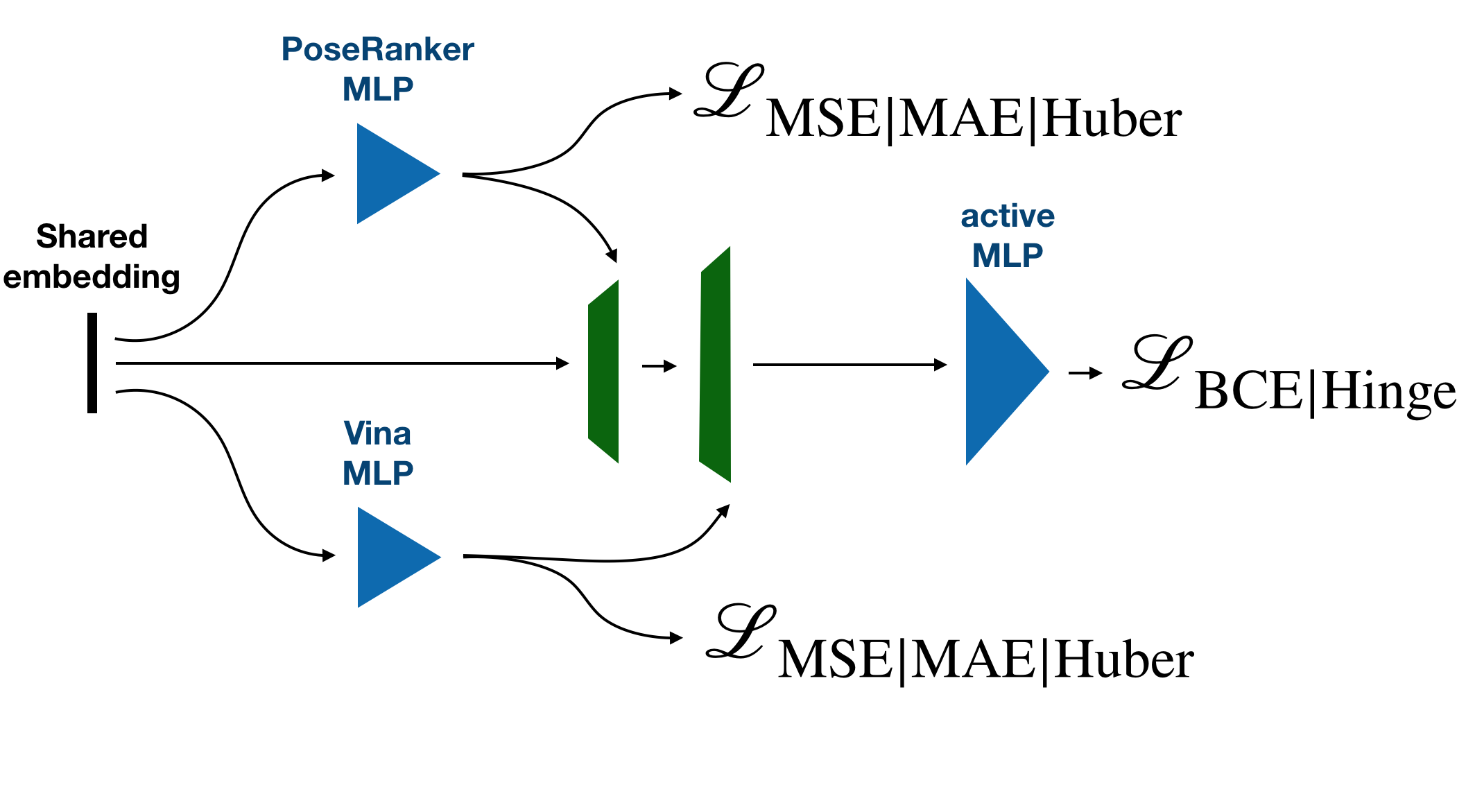}
  \caption{Active task conditioned on PoseRanker and Vina scores.}
  \label{mt_extra}
\end{figure}

The techniques used to condition network outputs on others (Figure~\ref{mt})
largely follow those outlined by Long et al.~\cite{long_conditional_2018}.
Given the shared embedding $\mathbf{z}_{\rm read} = g({f^{0...L}})$, the PoseRanker
score $\hat{y}_{\mathrm{pose}}$ and the Vina score $\hat{y}_{\mathrm{vina}}$
were computed by passing $\mathbf{z}_{\rm read}$ through separate MLPs
$y_{\mathrm{pose}}$ and $y_{\mathrm{vina}}$, respectively.
A conditioned embedding $\mathbf{z}'$ was then formed by passing
$\mathbf{z}$ through the conditional map:
\begin{equation*}
h : \mathbf{z}_{\rm read} \mapsto \begin{pmatrix}
 \mathbf{z}_{\rm read} \, \sigma(\hat{y}_{\mathrm{pose}}) \\
 \mathbf{z}_{\rm read} \left( 1 - \sigma(\hat{y}_{\mathrm{pose}}) \right)
\end{pmatrix}
\end{equation*}
where $\sigma : x \mapsto (1 + e^{-x})^{-1}$, and passed to a final MLP
$p_{\mathrm{active}}$ to compute the activity score.
Alternatively, $\mathbf{z}'$ may be passed through a second conditional
map,
\begin{equation*}
h' : \mathbf{z}' \mapsto \begin{pmatrix}
 \mathbf{z}' \\
 \hat{y}_{\mathrm{vina}}
\end{pmatrix}
\end{equation*}
to obtain an embedding $\mathbf{z}''$ that has been conditioned on both
the PoseRanker score and the Vina score (Figure~\ref{mt_extra}).

\subsection{Training}\label{model_training}
For every architecture, we trained 6 models for 10 epochs, each with \nicefrac{5}{6}th of the data as the training set, and \nicefrac{1}{6}th left out for cross-fold validation. Each data cross-fold contains clusters of proteins that share more than 70\% of the sequence similarity. We used the ADAM optimizer with the learning rate $\mathrm{lr} = 0.001$, and sampled targets with replacement, proportionally to the number of active compounds associated with that target. Targets without any measured active compounds were pruned from the training set.

\subsection{Data augmentation with pose-negatives}\label{pns}
We generated 64 ligand poses for every ligand-target pair with the CUina docking engine. Next, we reranked poses with PoseRanker according to their similarity to a native pose \cite{stafford_atomnet_2021}, and selected the top 16 poses. We used the top-ranked pose as a 'good' pose in training and scoring, whereas we used the bottom-ranked (16th) pose as a 'poor' (pose-negative) pose and considered inactive (non-binder).

\begin{figure}
  \centering
  \includegraphics[width=7cm]{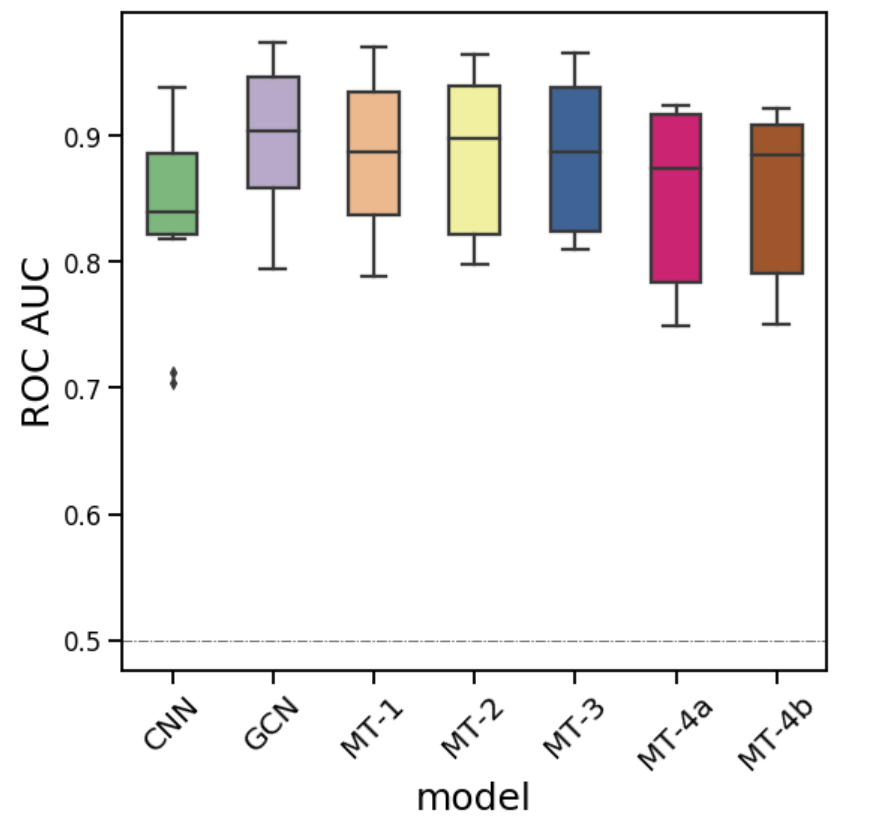}
  \caption{ ROC AUC performance on the D12 benchmark.}
  \label{d12}
\end{figure}

% \begin{figure}
%   \centering
%   \includegraphics[width=7cm]{Figures/Picasso.png}
%   \caption{Picasso problem experiment. $10^5$ diverse compounds (labeled as 0, non-binders) mixed with {\it c.a.} 300 kinase inhibitors (labeled as 1, binders) were docked and scored with at three binding sites i) ATP binding site ii) allosteric binding site iii) binding site at the SH2 domain, See Figure \ref{poses_zap70}b.}
%   \label{picasso}
% \end{figure}

\end{document}